\documentclass[a4paper,twocolumn,11pt,accepted=2021-08-20]{quantumarticle}
\pdfoutput=1
\usepackage[utf8]{inputenc}
\usepackage[english]{babel}
\usepackage[T1]{fontenc}
\usepackage{amsmath}
\usepackage{hyperref}
\usepackage{mathbbol}

\usepackage{tikz}
\usepackage{lipsum}

\usepackage[numbers,sort&compress]{natbib}

\begin{document}
\title{Reachability Deficits in Quantum Approximate Optimization of Graph Problems} 

\author{V.~Akshay}
\email{akshay.vishwanathan@skoltech.ru}
\homepage{http://quantum.skoltech.ru}
\orcid{0000-0002-5058-2585}
\affiliation{Skolkovo Institute of Science and Technology, 3 Nobel Street, Moscow, Russia 121205}

\author{H.~Philathong}
\orcid{0000-0002-1079-9341}
\affiliation{Skolkovo Institute of Science and Technology, 3 Nobel Street, Moscow, Russia 121205}

\author{I.~Zacharov}
\orcid{0000-0002-3256-6514}
\affiliation{Skolkovo Institute of Science and Technology, 3 Nobel Street, Moscow, Russia 121205}

\author{J.D.~Biamonte}
\orcid{0000-0002-0590-3327}
\affiliation{Skolkovo Institute of Science and Technology, 3 Nobel Street, Moscow, Russia 121205}

\maketitle
\begin{abstract}
  The quantum approximate optimization algorithm (QAOA) has become a cornerstone of contemporary quantum applications development. Here we show that the \emph{density} of problem constraints versus problem variables acts as a performance indicator.  Density is found to correlate strongly with approximation inefficiency for fixed depth QAOA applied to random graph minimization problem instances.  Further, the required depth for accurate QAOA solution to graph problem instances scales critically with density. Motivated by Google's recent experimental realization of QAOA, we preform a reanalysis of the reported data reproduced in an ideal noiseless setting. We found that the reported capabilities of instances addressed experimentally by Google, approach a rapid fall-off region in approximation quality experienced beyond intermediate-density. Our findings offer new insight into performance analysis of contemporary quantum optimization algorithms and contradict recent speculation regarding low-depth QAOA performance benefits.  
\end{abstract}

\section{Introduction}

Quantum approximate optimization (QAOA) is the most studied gate-based approach towards quantum enhanced optimization. Google's Sycamore quantum processor was recently used to demonstrate QAOA applied to graph minimization problems \cite{arute2020quantum}.  This recent study \cite{arute2020quantum} follows Sycamore's demonstration of computational capabilities surpassing those of classical supercomputers for specific sampling tasks \cite{arute2019quantum}.  

Such contemporary milestones have dramatically increased interest in harnessing quantum processors for more practical means.  Towards this goal, recent experimental demonstrations of quantum algorithms also include quantum chemistry \cite{aspuru2005simulated,o2016scalable}, machine learning \cite{biamonte2017quantum,xia2018quantum}, simulation of condensed matter systems \cite{lloyd1996universal,lamata2014efficient} as well as discrete optimization \cite{kadowaki1998quantum,arute2019quantum,pagano2019quantum}. While the ultimate prospects of quantum enhanced algorithms is tantalizing, the computational capacity of existing and near term applications remains unclear. 

Recently Google's Sycamore quantum processor, which is based on fifty four superconducting qubits, has demonstrated promising results related to quantum enhanced discrete optimization (via QAOA). The authors \cite{arute2020quantum} experimentally realized QAOA for three classes of graph optimization problems.  Such problems assign energy to graphs, which depend on nodes---represented as qubits---taking binary values. The experiments minimized this energy for three classes of graphs up to twenty three nodes \cite{arute2020quantum}. 

QAOA is an instance of the class of variational quantum algorithms.  Variational algorithms vary over a family of parameterized quantum states to minimize an objective function calculated from easy to measure observables. Proposed primarily in \cite{farhi2014quantum,mcclean2016theory}, the variational approach is in theory equivalent to the quantum circuit model under a polynomial (Karp) reduction \cite{biamonte2019universal}.  Hence, the variational approach is as powerful as a general quantum computer. However only restricted forms---which require significant classical resources in the optimization step---can currently be realized in practice.  

Indeed, the advantage of the variational approach is also its weakness.  A classical optimization loop iteratively adjusts a quantum state prepared by an adjustable quantum circuit (an ansatz).  High depth circuits of fixed structure have been proven to represent universal resources for quantum computation \cite{lloyd2018quantum,morales2019universality} and hence can emulate the gate model piece-wise. However, higher depth circuits demand a significant optimization task to be performed on a classical computer. In contrast, while reasonably short depth circuits have shown noise-free quantum advantage \cite{lin2016performance,crooks2018performance,marsh2019quantum,wang2018quantum, jiang2017near,morales2018variational}, less remains known about the depth of circuits required to enable practical advantage using noisy circuits.  

Changing to a state dependent picture and extending recent results related to satisfiability \cite{akshay2020reachability}, here we propose an order parameter which correlates with the performance of QAOA.  In particular, we consider the ratio of graph edges to graph nodes (called graph density).  Though a careful comparison, we empirically observe the following findings.~(i) Approximation inefficiency, characterised by metrics such as success probability, show an inverse dependence on density of randomly generated graph instances.~(ii) Longer depth circuits become a necessity for increasingly accurate approximations and this depth correlates with increasing density. Based on these results we, observe instances addressed in Google's experimental demonstration to represent a statistically narrow subset of random instances having densities that are at the edge of a fall-off region.

Contemporary demonstrations of QAOA serve as benchmarks to enable practically relevant demonstrations of optimization and other quantum algorithms \cite{otterbach2017unsupervised,willsch2020benchmarking,abrams2019implementation,bengtsson2019quantum,pagano2019quantum,qiang2018large}.  In this regard, our results illuminate the first QAOA performance indicators.

\section{Results} 

Google considered up to a depth-five QAOA ansatz on three separate classes of graph problems \cite{arute2020quantum}. Under experimental noise, Google was able to realize functional depth-three ansatz performance whereas depths four and five were critically limited by noise.  The depth three ansatz involves classical optimization over only six real parameters, three of which are constrained in the interval $[0, \pi)$ whereas the remaining three are in $[0, 2\pi)$.

\begin{figure*}[htp]
\minipage{\textwidth}
  \includegraphics[width=\textwidth]{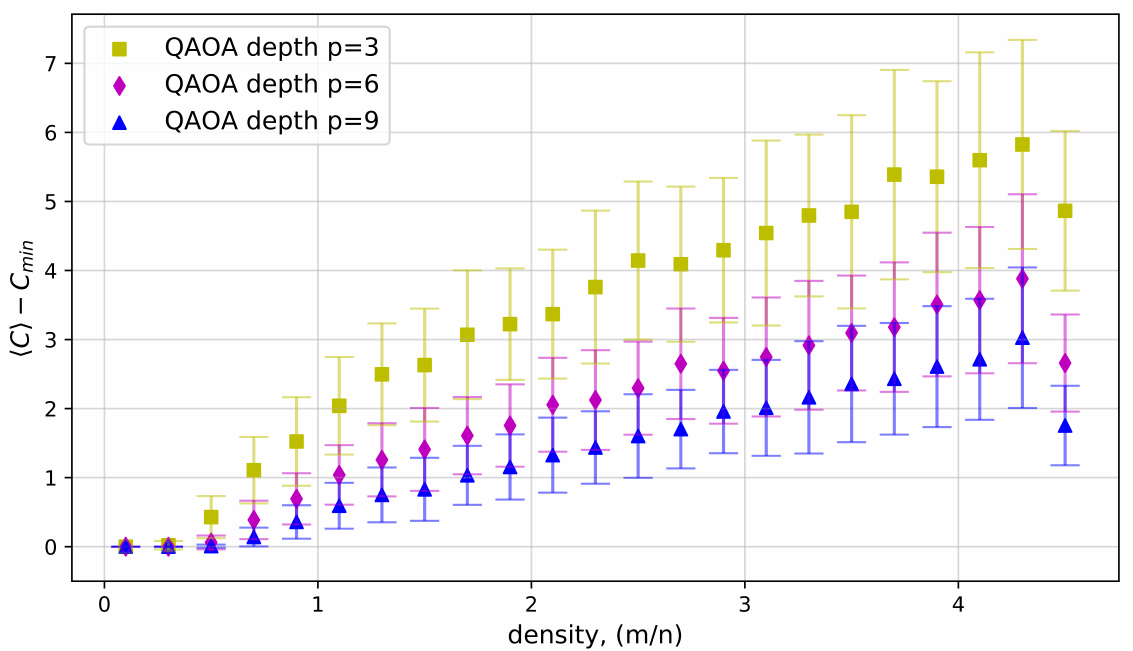}
  \includegraphics[width=\textwidth]{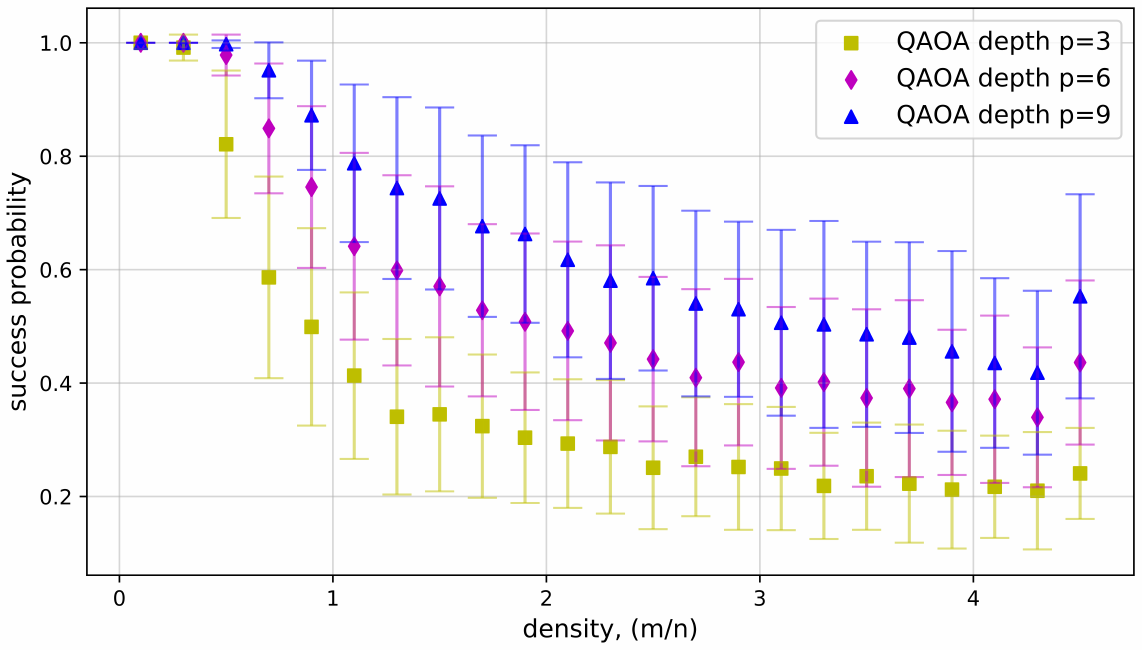}
\endminipage

\caption{QAOA performance metric as a function of density for depth $p=3,6,9$; (top) error in best possible approximation, and (bottom) success probability or ground state overlap. Each data point represents the average performance over statistics of $100$ uniform random graph instances, $G_{n,m}$ with nodes $n=10$. Here we see the strong correlation between approximation inefficiency and density in both performance metrics. }
\label{fixed_n_performance}
\end{figure*}

We reproduced Google's ideal (noiseless) data using exact numerical emulations (see Fig.~\ref{g_metric}). Further, we show that QAOA on random graph instances generated at varying densities exhibit strong density dependent performance. Although higher depths achieve better performance, the density dependence is still exhibited (see Fig.~\ref{fixed_n_performance}). Importantly, higher depths necessitate increasingly more parameters to be optimized in the classical step. 

\begin{figure*}[htp]
\minipage{\textwidth}
  \includegraphics[width=\textwidth]{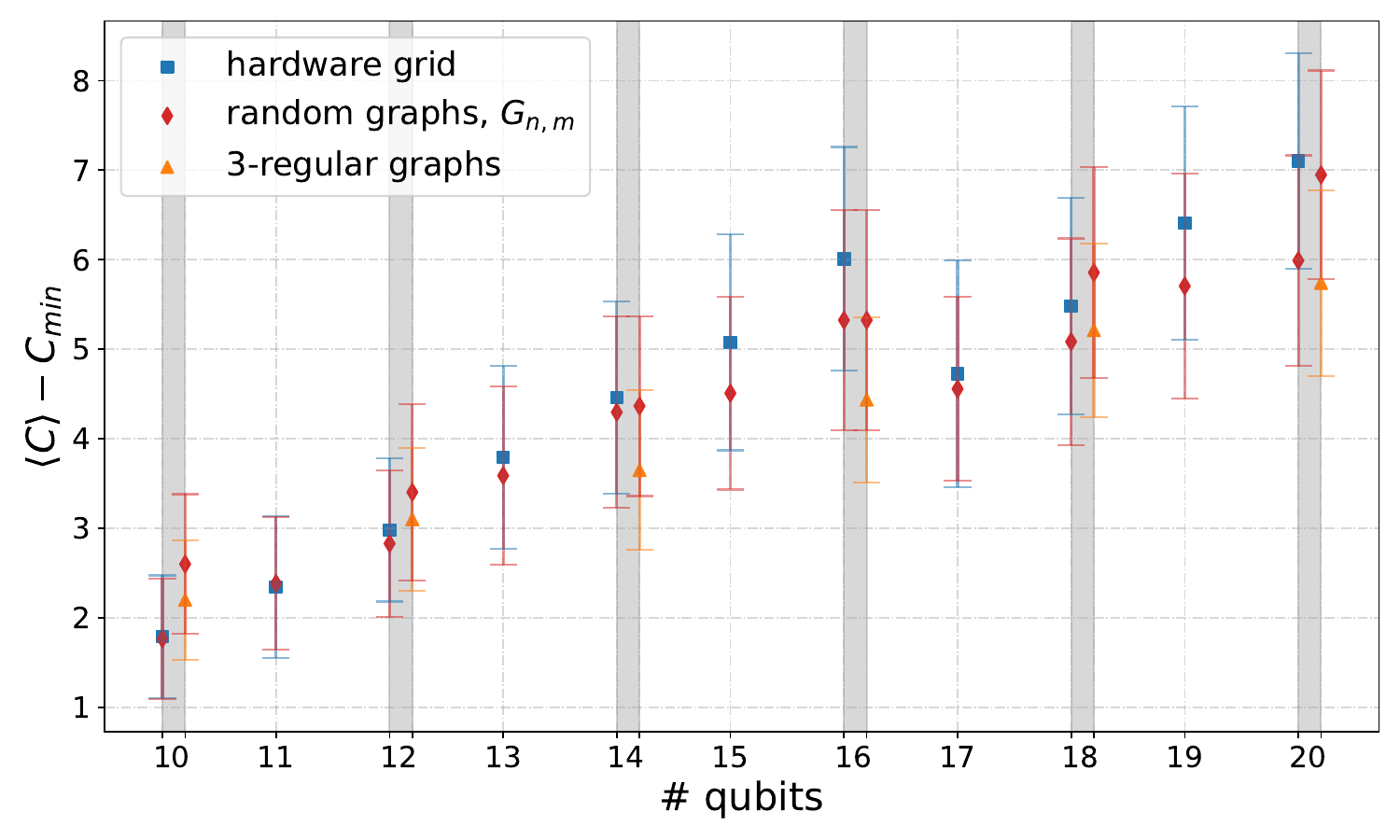}
  \includegraphics[width=\textwidth]{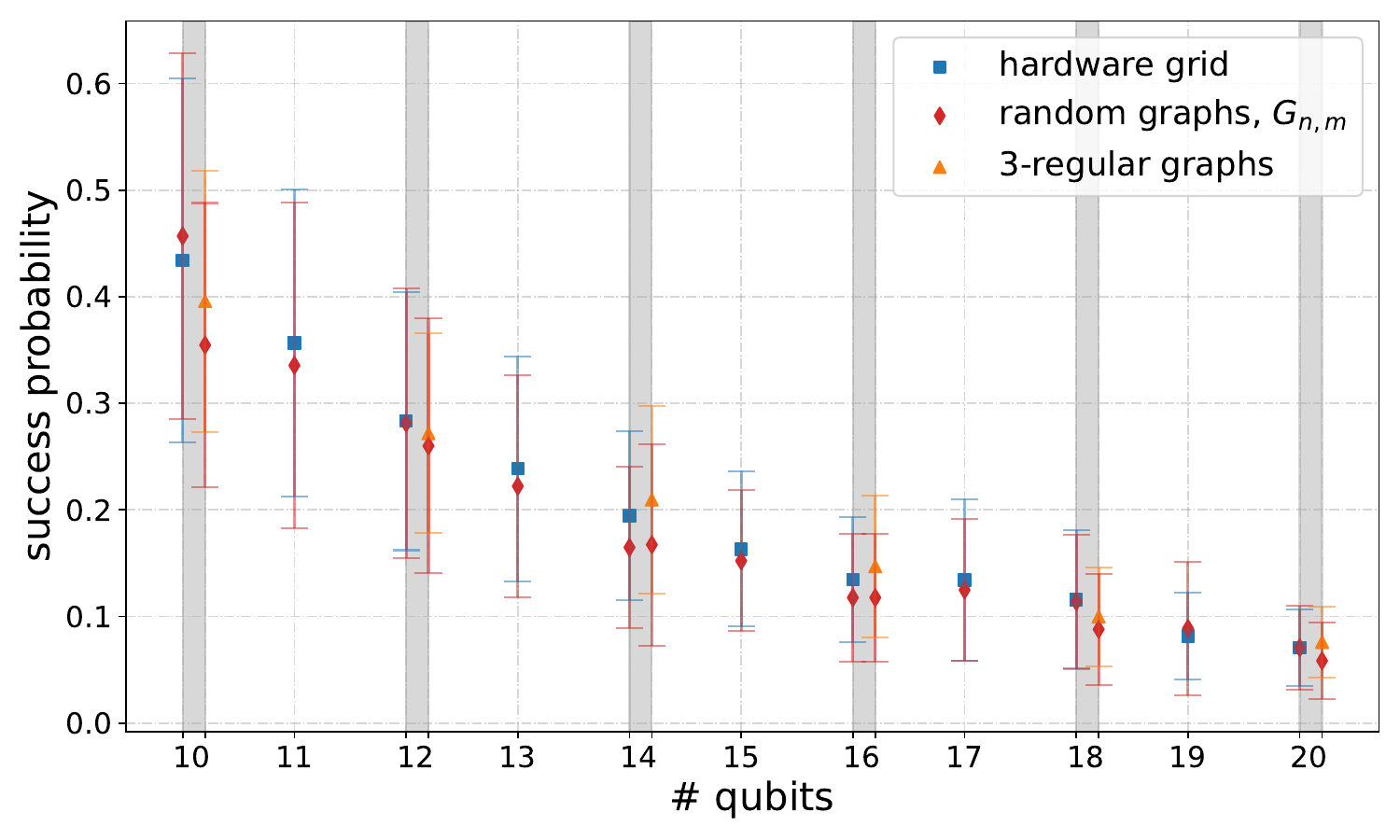}
\endminipage

\caption{Comparing (top) error in best possible approximation and (bottom) success probability of the three graph families. For even numbers of qubits, the left vertical pair represents the grid to random graph comparison, and the right vertical pair represents the 3-regular to random graph comparison. For odd numbers of qubits, the grid to random graph comparison is shown (since 3-regular graphs can be generated only with even number of nodes). Each data point represents the average performance of depth $p=3$ QAOA over statistics of $100$ randomly generated instances. By comparison, we do not observe any topology related bias in QAOA performance that can be of statistical relevance.} 
\label{comparision_performance}
\end{figure*}

Based on the comparison of random graphs at approximately the same densities with the restrictive graph instances Google considered, we do not observe any  topology  dependent  performance  bias  that can  be  attributed statistical relevance. This finding suggests density---albeit a coarse graining---is the salient limiting performance indicator for fixed depth QAOA (see Fig.~\ref{comparision_performance}). 

\begin{figure*}[htp]
\minipage{\textwidth}
    \centering
  \includegraphics[width=0.80\textwidth]{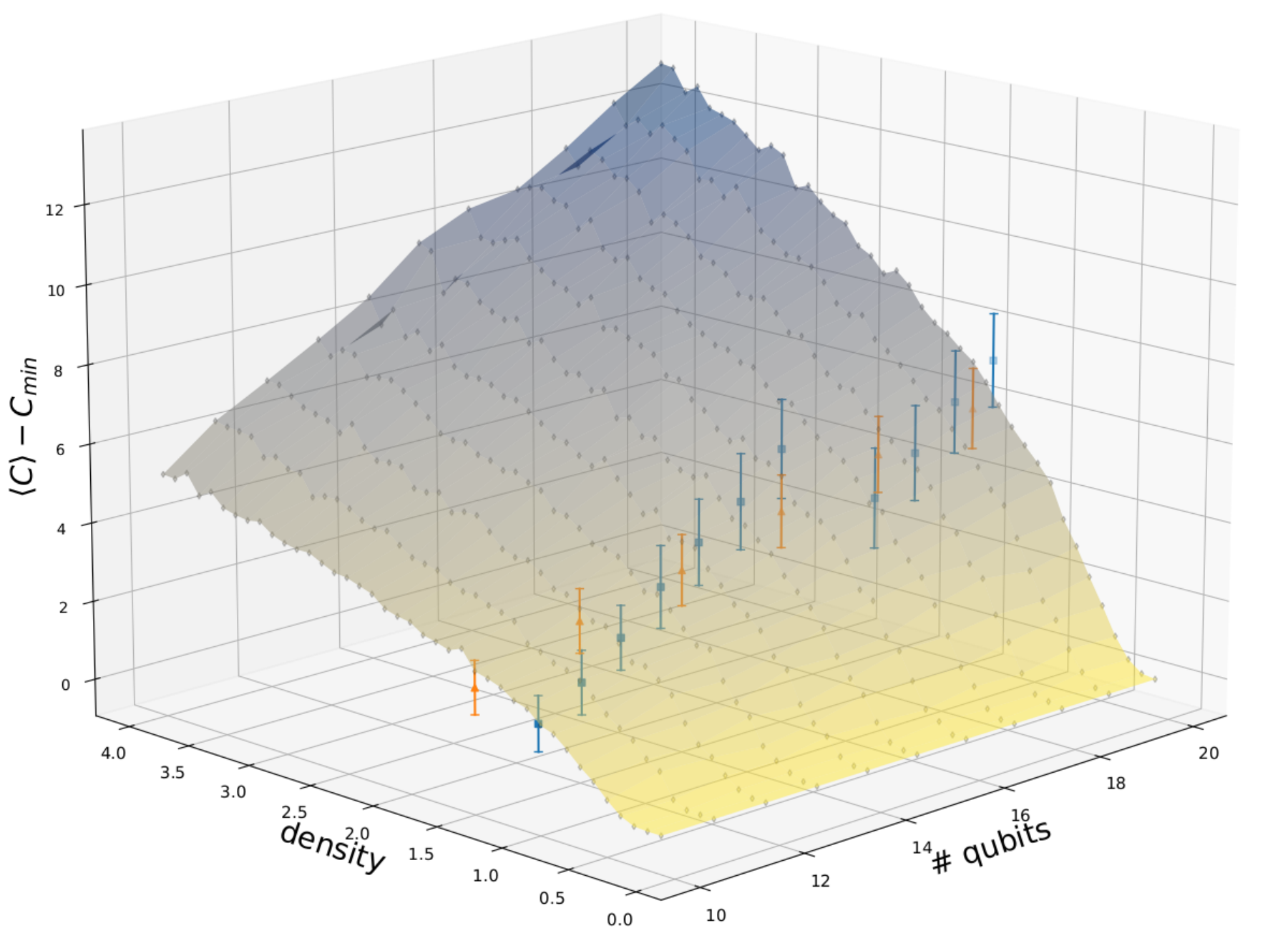}
  \includegraphics[width=0.80\textwidth]{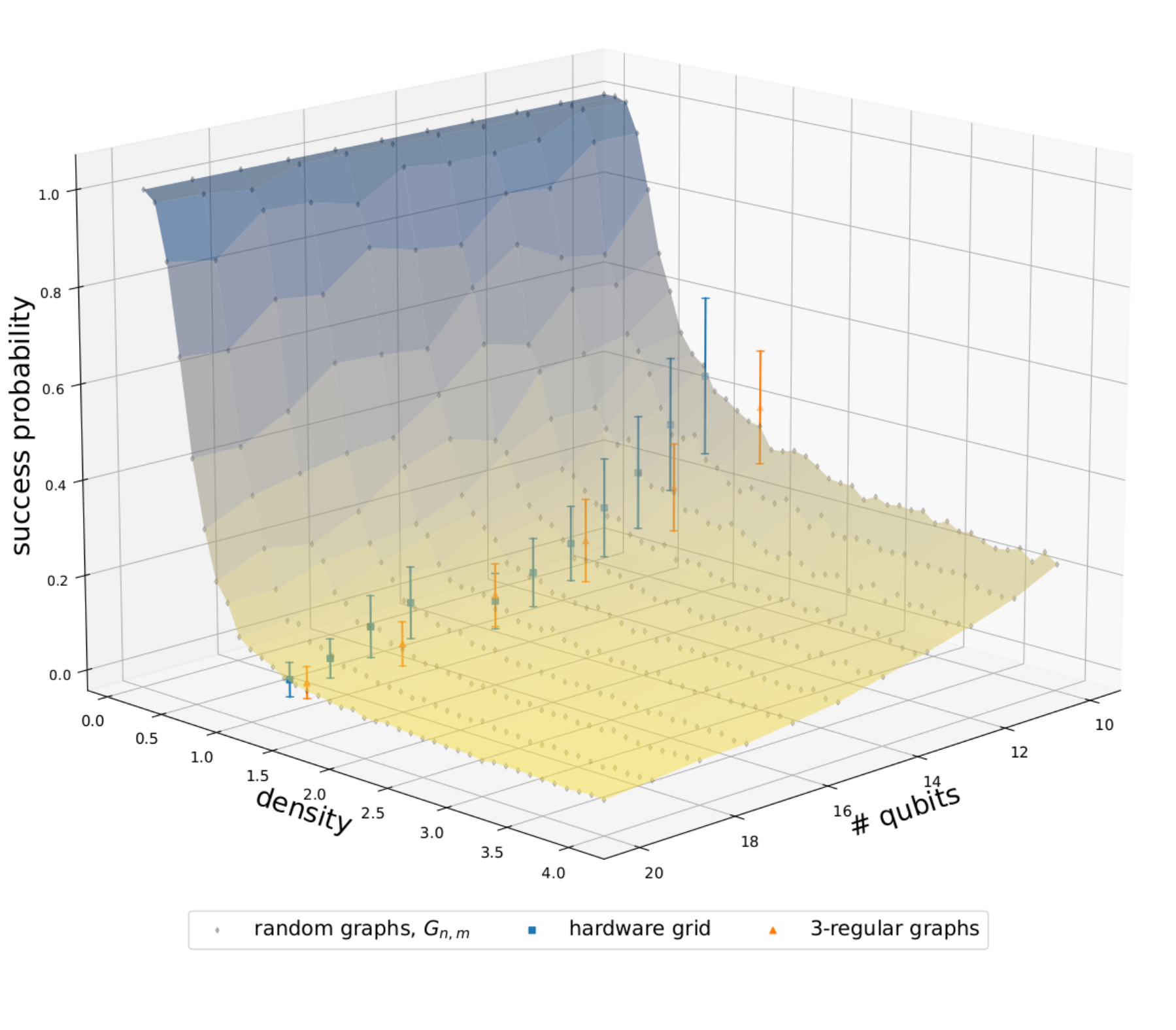}
\endminipage

\caption{3D landscape of depth--$3$ QAOA performances; (top) error in best possible approximation and (bottom) success probability. Google data points are marked on the landscape according to the mean density and performance of the considered instances. We observe a sharp fall-off behaviour in QAOA performance which appears to saturate logistically beyond intermediate density. We find that Google data points lie between densities $1$ and $1.5$, where a low depth QAOA ansatz can approximate a solution(s) efficiently.}
\label{performance_3dlandscape}
\end{figure*}

\begin{figure*}[htp]
\minipage{\textwidth}
  \includegraphics[width=0.95\textwidth]{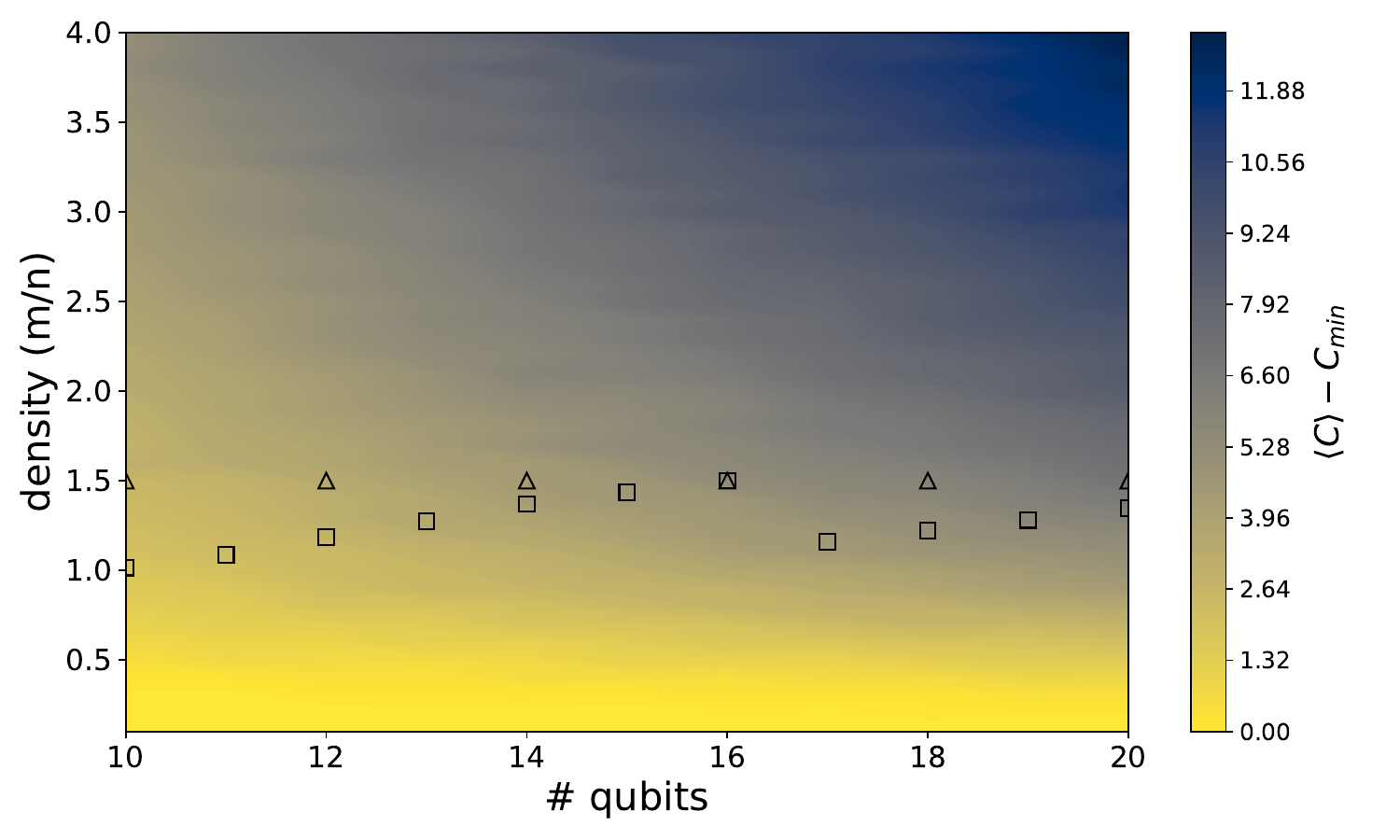}
  \includegraphics[width=0.95\textwidth]{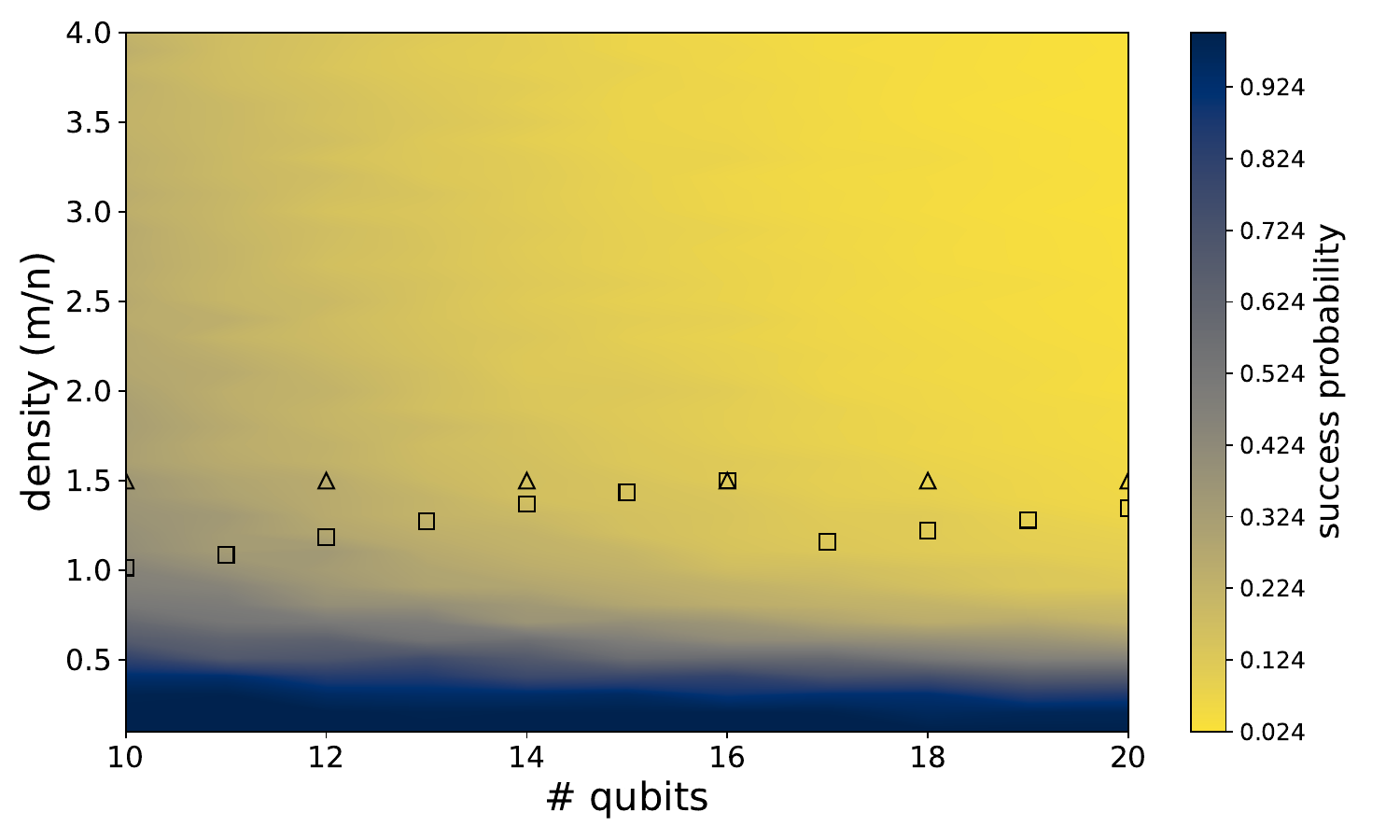}
\endminipage

\caption{2D projection of depth--$3$ QAOA performance; (top) error in best possible approximation and (bottom) success probability. Google data points are marked on the landscape according to the mean density of the considered instances. Squares represent grid instances and Triangles represent 3 regular graphs. Google data points lie between density $1$ and $1.5$, where a low depth QAOA ansatz can approximate a solution(s) efficiently.}
\label{performance_2D projection}
\end{figure*}

With density as an order parameter, we find that solutions of random instances generated below densities of $0.5$ to show the best performance independent of the number of nodes. Beyond density $0.5$, we find the performance to strongly depend on density for fixed number of nodes. In particular, we observe a sharp fall-off behaviour in performance which appears to saturate logistically beyond density $1.5$. It is under this new outlook Google's instances are compared and contrasted (see Fig.~\ref{performance_3dlandscape} and Fig.~\ref{performance_2D projection}).

Firstly, the authors~\cite{arute2020quantum} consider the Hardware-Grid problem wherein the generated random instances are native to the inherent hardware connectivity layout of Google's Sycamore chip (sometimes called, \emph{interaction graph}). Here the instances generated by varying the number of nodes ($n$) which we found to appear at densities between $1.0$ and $1.5$. Our exact simulation of depth three QAOA suggests the success probability (or the state overlap) to worsen with the number of nodes (increasing density), achieving a meager $0.07$ for nodes $n=20$ (at density = 1.3). 

Secondly, the experiments considered $3$-Regular Graphs. In this case the randomly generated instances have an even number of nodes with constant density $1.5$. We found success probability to worsen with the number of nodes achieving roughly the same value of $0.07$ for nodes $n=20$ (at density = 1.5).

In both cases we find the considered instances appear on the edge of a fall-off region in our analysis of the density dependent performance landscape. It is for instances beyond these densities (that is, beyond critical approximation density fall off) that we observe limited performance for depth three QAOA ansatz.  

Finally, the mean field Sherrington-Kirkpatrick (SK) model is considered where the generated graph instances are fully connected and have the maximum possible density ($(n-1)/2$) for $n$ nodes ranging from $10$ to $20$. 

Unlike the previous two cases, the SK instances scale up density linearly with $n$. Furthermore, for any fixed $n$, such graph instances have the highest density and therefore QAOA should indicate poor performance. This agrees exactly with our emulation. For each $n$, we observe the success probability for the SK model to be the lowest in the considered family of graphs (see the bottom pane on Fig.~\ref{g_metric}).

\section{Methods}

\subsection{QAOA} Given an objective function which encodes the solution to an optimization problem in the ground state of a Hamiltonian, $\mathcal{V}$, the usual QAOA procedure is as follows,
\begin{enumerate}
        \item Generate ansatz states, $|\psi(\boldsymbol{\gamma},\boldsymbol{\beta})\rangle$ on a quantum computer, where $\boldsymbol{\gamma}=(\gamma_{1},\gamma_{2},\cdots,\gamma_{p})$ and $\boldsymbol{\beta}=(\beta_{1},\beta_{2},\cdots,\beta_{p})$ are tunable real parameters over some fixed range. The state is prepared by applying a sequence of 2\textit{p}-parameter unitary gates acting on the reference state $|+\rangle^{\otimes{n}}$ which is the symmetric superposition of all $2^{n}$ computational basis states as follows, 
    \begin{equation}\label{qaoa equation}
        |\psi(\boldsymbol{\gamma},\boldsymbol{\beta})\rangle=\prod_{k=1}^{p} \mathcal{U}(\gamma_{k},\beta_{k})|+\rangle^{\otimes{n}},
    \end{equation} 
    where
    \begin{equation}\label{driverandproblem}
        \mathcal{U}(\gamma_{k},\beta_{k})=\exp(-i \beta_{k}\mathcal{H}_{x}) \cdot \exp(-i \gamma_{k}\mathcal{V}).
    \end{equation}
    The Hamiltonian $\mathcal{H}_{x}=\sum_{i} \sigma_{x}^{(i)}$ is called the driver Hamiltonian, and $\mathcal{V}$ is the problem Hamiltonian here ground states correspond to the bit strings that minimizes the objective function of interest.
    \item Measurement of the state in Eq.~\eqref{qaoa equation} is done to compute the expected value, $\langle \psi(\boldsymbol{\gamma},\boldsymbol{\beta})|\mathcal{V}|\psi(\boldsymbol{\gamma},\boldsymbol{\beta})\rangle$. This is an approximation which can be calculated efficiently on a classical computer.
    \item Steps 1 and 2 are repeated and a classical optimization algorithms is used to assign a set of optimum parameters, $\boldsymbol{\gamma}^{*}$ and $\boldsymbol{\beta}^{*}$ that minimize  $\langle \psi(\boldsymbol{\gamma},\boldsymbol{\beta})|\mathcal{V}|\psi(\boldsymbol{\gamma},\boldsymbol{\beta})\rangle$.  $\langle \psi(\boldsymbol{\gamma}^{*},\boldsymbol{\beta}^{*})|\mathcal{V}|\psi(\boldsymbol{\gamma}^{*},\boldsymbol{\beta}^{*})\rangle$ then represents an approximate solution to the optimization problem.
\end{enumerate}

\subsection{Problem description}

In this paper we study minimization problem on graphs.  The objective function of interest can be defined as follows:  

Given an undirected weighted graph $G=(V,E)$, with $|V| = n$ nodes and $|E| = m$ edges. $E$ denotes the set of edges in $G$ as  $E = \lbrace (i,j) \rbrace$, where $i \neq j $ and $i,j \in V$. In general, the edge weights are defined as a map $w:\, E \rightarrow \mathbb{R}$.  Yet in our case and without loss of generality, we restrict to sampling $w_{ij}$s from a uniform distribution over $\lbrace -1,+1 \rbrace$. 

Minimization problems on $G$ are defined thorough a cost-function $\mathcal{C}(\boldsymbol{z})$ with a corresponding embedding as quantum Hamiltonian operator. Here we employ the following as our objective function,
\begin{equation}
    \mathcal{C} = \sum_{(i,j) \in E} w_{ij}Z_{i}Z_{j},
\end{equation}
where $\boldsymbol{z}$ denotes the bitstring or bit assignment to the nodes in $G$ and $Z_{i}$ corresponds to the Pauli $Z$ operator action on the $i^{th}$ qubit.
In general, with appropriate cost-function construction and Ising spin encoding, minimization on graphs can represent solving a family of NP-hard optimization problems such as MAX-CUT, Vertex cover and Maximum Independent Set to name a few.

\subsection{Performance metrics} 

QAOA performance is analysed under the following two performance metrics, 
\begin{enumerate}
    \item the error in best possible approximation,
    \item the success probability or the ground state overlap.
\end{enumerate}

Let $|\psi\rangle$, be the \textit{p}--depth QAOA ansatz states generated by Eq.~\eqref{qaoa equation} and let ${\mathcal A}$ be a subset that is not nessisarily equal to $\mathbb{C}_{2}^{\otimes n}$.  Then we define the error in best possible approximation as 
\begin{equation}\label{reachabilitydef}
    f = \min_{|\psi\rangle \in {\mathcal R}\subset \mathbb{C}_{2}^{\otimes n}}\langle\psi|\mathcal{C}|\psi\rangle - \min_{|\phi\rangle \in \mathbb{C}_{2}^{\otimes n}} \langle\phi|\mathcal{C}|\phi\rangle.
\end{equation}
This quantity characterises the limiting performance of QAOA. The first term on the right indicates minimization over the set of reachable states generated by a fixed depth QAOA ansatz whereas the second term represents the exact optimum value of $\mathcal{C}$ computed by minimization over the entire Hilbert space.

To calculate the success probability or the ground state overlap, let $\lbrace |\text{gs}_{i}\rangle\rbrace$ be the set of \textit{d} degenerate ground states of $\mathcal{C}$ then the overlap is given by,
\begin{equation}
    \eta = \sum_{i=1}^{d}{ |\langle\psi_{p}(\boldsymbol{\gamma},\boldsymbol{\beta})|\text{gs}_{i} \rangle |^{2}}.
\end{equation}

A well known performance metric is the approximation ratio defined as,
\begin{equation}
    r = \frac{\langle \mathcal{C} \rangle}{\mathcal{C}_{min}}.
\end{equation}
It is under this performance metric that Google's data are represented \cite{arute2020quantum}. Although this metric is bounded between $[0,1]$, and it can directly be used to contrast performance with other polynomial-time approximation algorithms, we observe the effect of density dependence to be visually suppressed. This is due to the logistic behaviour of the error in best possible approximation, $\langle \mathcal{C} \rangle - \mathcal{C}_{min}$. Since both $\langle \mathcal{C} \rangle$ and $\mathcal{C}_{min}$ grow with density for any given $n$ and fixed depth QAOA ansatz, the ratio visually conceals density dependence. This is further illustrated when we consider success probability as our performance indicator. Under this choice we clearly observe the limiting performance of QAOA.

\subsection{Graph Instances} 

Here we present the construction of random problem instances used in our work. 

\emph{Fixed $n$ case.} We first show the approximation inefficiency or the feature of density dependent performance under the two metrics by studying QAOA on the uniform random graph model, $G_{n,m}$ for fixed $n$. The density in the case of such graph problems can be defined as the ratio of the number of edges to the number of nodes, $m/n$.

Random instances are generated by initializing an empty graph on the vertex set of size $|V|= n$. Then, $m$ edges are constructed in such a way that all possible $\binom{\binom{n}{2}}{m}$ choices are equally likely with random weights drawn from $\lbrace -1,+1 \rbrace$.

\emph{3-regular graphs}. A 3-regular graph, also known as a cubic graph, is when each node in the graph has degree equal to 3. The density of generated instance in this case is constant ($density = 1.5$) as the number of edges required for constructing a 3-regular graph is given by $3\times\frac{n}{2}$, for even $n$.  

\emph{2-D grid graphs}. These graphs are interesting mainly due to the inherent hardware connectivity of Google's Sycamore. Such graphs are generated by constructing an $N \times N$ square grid and randomly deleting nodes on the boundaries to construct random grid instance with $n \leq N^2$ nodes.
Unlike the case of regular graphs, 2-D grid graphs do not share a constant density across the generated instance. The maximum number of edges in a square/partially square grid is given by $\lfloor 2n - 2\sqrt{n} \rfloor$. Therefore, at most density of such instances is bounded above by $2.0$.

\subsection{Numerical Methods} 

The large calculations were performed on Skotlech's Zhores computer \cite{Zacharov2019}. Our simulation of QAOA is based on recovering the exact state vector in a noiseless setting. Therefore in calculation of success probability infinite sampling is assumed. Standard L-BFGS-B optimizer from SciPy with 25 random seed runs are used to find good optimal parameters to minimize the objective function. The procedure is then repeated over 100 randomly generated graph instances at each respective density to evaluate average performance and the corresponding standard deviation (represented in the plots as error-bars). The code is written in Python with Intel optimised libraries and is available on reasonable request.

\section{Conclusions}

\begin{figure*}[htp]
\minipage{\textwidth}
    
  \includegraphics[width = \textwidth]{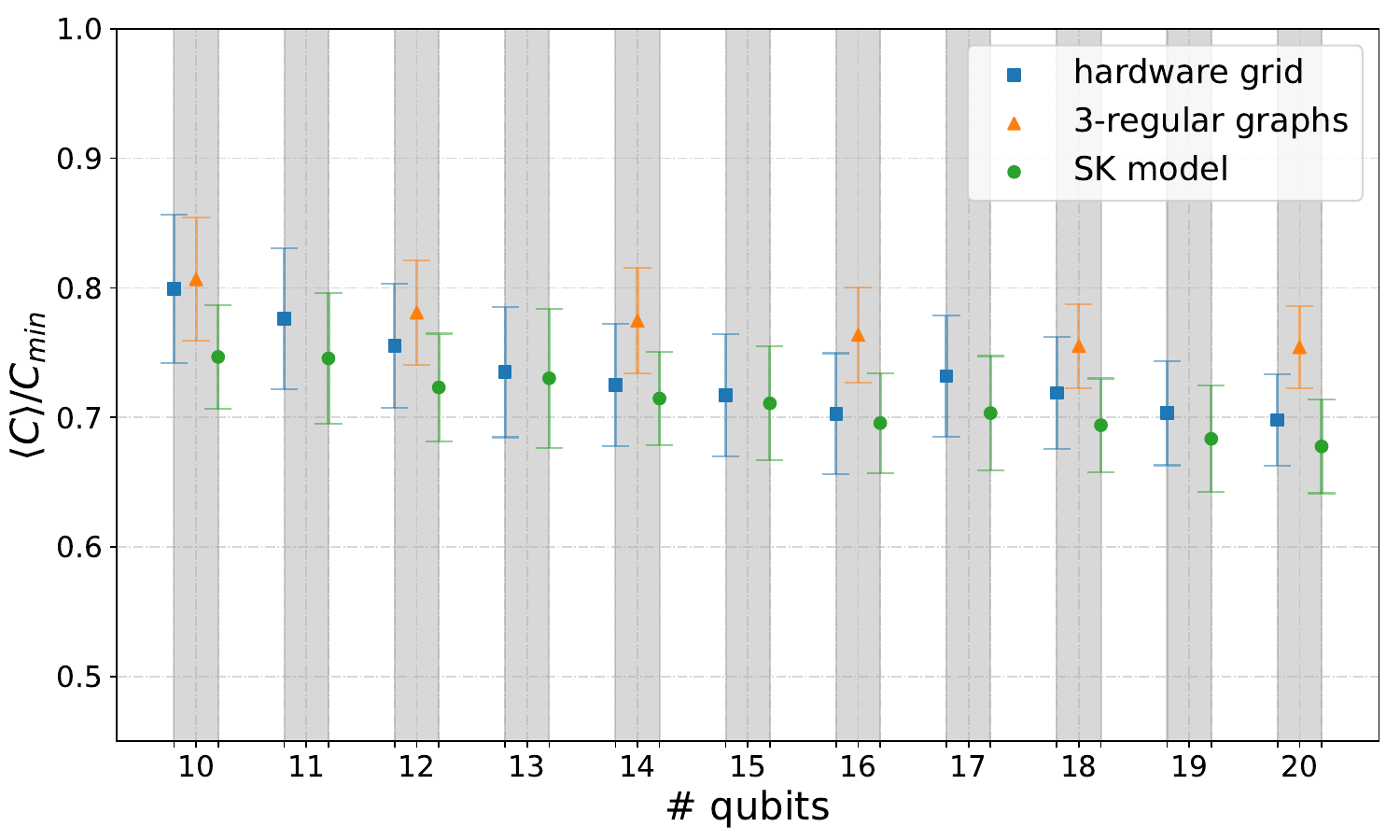}
  \includegraphics[width = \textwidth]{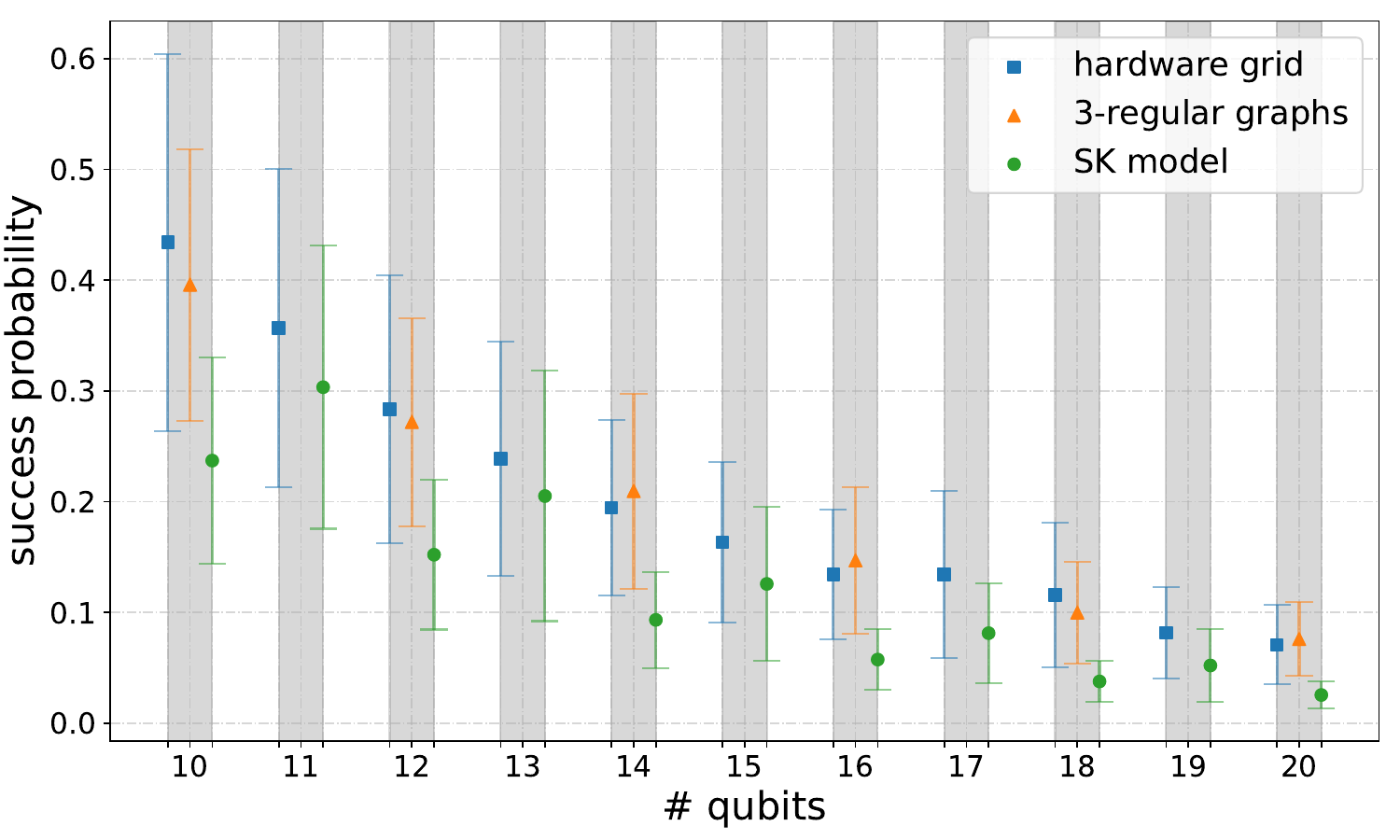}
\caption{Reproducing Google's ideal noiseless data for the three families of graph, (i) hardware grid graphs (blue), (ii) 3-regular graphs (orange), and (iii) SK model or complete graphs (green). Each data point represents the average performance of depth $p=3$ QAOA over statistics of $100$ randomly generated instances. We observe that the effect of density dependence in QAOA performance is clearly observed when considering the success probability as the metric (bottom). On other hand, it remains visually suppressed using Google's approximation ratio (top).}
\label{g_metric}  
\endminipage
\end{figure*}

To assess QAOA performance, Google applies {\it approximation ratio}---a standard technique in the classical theory of graph optimization (see e.g.~\cite{arute2020quantum}). Under this setting, an almost constant performance across the three graph problem families is visually observed---reproduced in Fig.~\ref{g_metric} top pane.  

It is only under other indicators such as (i) best approximation error and (ii) success probability---see Methods---that the density dependence in approximation inefficiency is visually revealed (a.k.a.~reachability deficits). Although the effect can be inferred from Google's approximation ratio, the effect of reachability deficits are highly suppressed visually.   

We finally assert that future experimental and numerical studies related to QAOA should report and contrast algorithmic performance with instance densities.

\begin{acknowledgments}
The authors acknowledge support from the project, {\it Leading Research Center on Quantum Computing} (Agreement No.~014/20). The authors acknowledge the use of Skoltech's Zhores supercomputer for obtaining the numerical results presented in this paper.

{\it Competing interests.} The authors declare no competing interests.

{\it Author contributions.} All authors conceived and developed the theory and design of this study and verified the methods. JB proposed the study and supervised its execution.  VA, HP and IZ developed and deployed the parallel code to collect numerical data.  All authors contributed to interpret the results and wrote the manuscript.

{\it Data and code availability.} The data that supports the findings of this study are available within the article. The code for generating the data is available on GitHub at \hyperlink{https://github.com/vishwanathanakshay/QAOA}{https://github.com/vishwanathanakshay/QAOA}. 
\end{acknowledgments}

\bibliographystyle{unsrtnat}
\bibliography{mybibliography}

\begin{thebibliography}{28}
\providecommand{\natexlab}[1]{#1}
\providecommand{\url}[1]{\texttt{#1}}
\expandafter\ifx\csname urlstyle\endcsname\relax
  \providecommand{\doi}[1]{doi: #1}\else
  \providecommand{\doi}{doi: \begingroup \urlstyle{rm}\Url}\fi

\bibitem[Harrigan et~al.(2021)Harrigan, Sung, Neeley, Satzinger, Arute, Arya,
  Atalaya, Bardin, Barends, Boixo, et~al.]{arute2020quantum}
Matthew~P Harrigan, Kevin~J Sung, Matthew Neeley, Kevin~J Satzinger, Frank
  Arute, Kunal Arya, Juan Atalaya, Joseph~C Bardin, Rami Barends, Sergio Boixo,
  et~al.
\newblock Quantum approximate optimization of non-planar graph problems on a
  planar superconducting processor.
\newblock \emph{Nature Physics}, 17\penalty0 (3):\penalty0 332--336, 2021.
\newblock \doi{10.1038/s41567-020-01105-y}.

\bibitem[Arute et~al.(2019)Arute, Arya, Babbush, Bacon, Bardin, Barends,
  Biswas, Boixo, Brandao, Buell, et~al.]{arute2019quantum}
Frank Arute, Kunal Arya, Ryan Babbush, Dave Bacon, Joseph~C Bardin, Rami
  Barends, Rupak Biswas, Sergio Boixo, Fernando~GSL Brandao, David~A Buell,
  et~al.
\newblock Quantum supremacy using a programmable superconducting processor.
\newblock \emph{Nature}, 574\penalty0 (7779):\penalty0 505--510, 2019.
\newblock \doi{10.1038/s41586-019-1666-5}.

\bibitem[Aspuru-Guzik et~al.(2005)Aspuru-Guzik, Dutoi, Love, and
  Head-Gordon]{aspuru2005simulated}
Al{\'a}n Aspuru-Guzik, Anthony~D Dutoi, Peter~J Love, and Martin Head-Gordon.
\newblock Simulated quantum computation of molecular energies.
\newblock \emph{Science}, 309\penalty0 (5741):\penalty0 1704--1707, 2005.
\newblock \doi{10.1126/science.1113479}.

\bibitem[O’Malley et~al.(2016)O’Malley, Babbush, Kivlichan, Romero,
  McClean, Barends, Kelly, Roushan, Tranter, Ding, et~al.]{o2016scalable}
Peter~JJ O’Malley, Ryan Babbush, Ian~D Kivlichan, Jonathan Romero, Jarrod~R
  McClean, Rami Barends, Julian Kelly, Pedram Roushan, Andrew Tranter, Nan
  Ding, et~al.
\newblock Scalable quantum simulation of molecular energies.
\newblock \emph{Physical Review X}, 6\penalty0 (3):\penalty0 031007, 2016.
\newblock \doi{10.1103/PhysRevX.6.031007}.

\bibitem[Biamonte et~al.(2017)Biamonte, Wittek, Pancotti, Rebentrost, Wiebe,
  and Lloyd]{biamonte2017quantum}
Jacob Biamonte, Peter Wittek, Nicola Pancotti, Patrick Rebentrost, Nathan
  Wiebe, and Seth Lloyd.
\newblock Quantum machine learning.
\newblock \emph{Nature}, 549\penalty0 (7671):\penalty0 195--202, 2017.
\newblock \doi{10.1038/nature23474}.

\bibitem[Xia and Kais(2018)]{xia2018quantum}
Rongxin Xia and Sabre Kais.
\newblock Quantum machine learning for electronic structure calculations.
\newblock \emph{Nature communications}, 9\penalty0 (1):\penalty0 1--6, 2018.
\newblock \doi{10.1038/s41467-018-06598-z}.

\bibitem[Lloyd(1996)]{lloyd1996universal}
Seth Lloyd.
\newblock Universal quantum simulators.
\newblock \emph{Science}, pages 1073--1078, 1996.
\newblock URL \url{https://www.jstor.org/stable/2899535}.

\bibitem[Lamata et~al.(2014)Lamata, Mezzacapo, Casanova, and
  Solano]{lamata2014efficient}
Lucas Lamata, Antonio Mezzacapo, Jorge Casanova, and Enrique Solano.
\newblock Efficient quantum simulation of fermionic and bosonic models in
  trapped ions.
\newblock \emph{EPJ Quantum Technology}, 1\penalty0 (1):\penalty0 9, 2014.
\newblock \doi{10.1140/epjqt9}.

\bibitem[Kadowaki and Nishimori(1998)]{kadowaki1998quantum}
Tadashi Kadowaki and Hidetoshi Nishimori.
\newblock Quantum annealing in the transverse ising model.
\newblock \emph{Physical Review E}, 58\penalty0 (5):\penalty0 5355, 1998.
\newblock \doi{10.1103/PhysRevE.58.5355}.

\bibitem[Pagano et~al.(2020)Pagano, Bapat, Becker, Collins, De, Hess, Kaplan,
  Kyprianidis, Tan, Baldwin, et~al.]{pagano2019quantum}
Guido Pagano, Aniruddha Bapat, Patrick Becker, Katherine~S Collins, Arinjoy De,
  Paul~W Hess, Harvey~B Kaplan, Antonis Kyprianidis, Wen~Lin Tan, Christopher
  Baldwin, et~al.
\newblock Quantum approximate optimization of the long-range ising model with a
  trapped-ion quantum simulator.
\newblock \emph{Proceedings of the National Academy of Sciences}, 117\penalty0
  (41):\penalty0 25396--25401, 2020.
\newblock \doi{10.1073/pnas.2006373117}.

\bibitem[Farhi et~al.(2014)Farhi, Goldstone, and Gutmann]{farhi2014quantum}
Edward Farhi, Jeffrey Goldstone, and Sam Gutmann.
\newblock A quantum approximate optimization algorithm.
\newblock \emph{arXiv preprint arXiv:1411.4028}, 2014.
\newblock URL \url{https://arxiv.org/abs/1411.4028}.

\bibitem[McClean et~al.(2016)McClean, Romero, Babbush, and
  Aspuru-Guzik]{mcclean2016theory}
Jarrod~R McClean, Jonathan Romero, Ryan Babbush, and Al{\'a}n Aspuru-Guzik.
\newblock The theory of variational hybrid quantum-classical algorithms.
\newblock \emph{New Journal of Physics}, 18\penalty0 (2):\penalty0 023023,
  2016.
\newblock \doi{10.1088/1367-2630/18/2/023023}.

\bibitem[Biamonte(2021)]{biamonte2019universal}
Jacob Biamonte.
\newblock Universal variational quantum computation.
\newblock \emph{Phys. Rev. A}, 103:\penalty0 L030401, Mar 2021.
\newblock \doi{10.1103/PhysRevA.103.L030401}.

\bibitem[Lloyd(2018)]{lloyd2018quantum}
Seth Lloyd.
\newblock Quantum approximate optimization is computationally universal.
\newblock \emph{arXiv preprint arXiv:1812.11075}, 2018.
\newblock URL \url{https://arxiv.org/abs/1812.11075}.

\bibitem[Morales et~al.(2020)Morales, Biamonte, and
  Zimbor{\'a}s]{morales2019universality}
Mauro~ES Morales, Jacob~D Biamonte, and Zolt{\'a}n Zimbor{\'a}s.
\newblock On the universality of the quantum approximate optimization
  algorithm.
\newblock \emph{Quantum Information Processing}, 19\penalty0 (9):\penalty0
  1--26, 2020.
\newblock \doi{10.1007/s11128-020-02748-9}.

\bibitem[{Yen-Yu Lin} and {Zhu}(2016)]{lin2016performance}
Cedric {Yen-Yu Lin} and Yechao {Zhu}.
\newblock {Performance of QAOA on Typical Instances of Constraint Satisfaction
  Problems with Bounded Degree}.
\newblock \emph{arXiv e-prints}, art. arXiv:1601.01744, January 2016.
\newblock URL \url{https://arxiv.org/abs/1601.01744}.

\bibitem[Crooks(2018)]{crooks2018performance}
Gavin~E Crooks.
\newblock Performance of the quantum approximate optimization algorithm on the
  maximum cut problem.
\newblock \emph{arXiv preprint arXiv:1811.08419}, 2018.
\newblock URL \url{https://arxiv.org/abs/1811.08419}.

\bibitem[Marsh and Wang(2019)]{marsh2019quantum}
Samuel Marsh and JB~Wang.
\newblock A quantum walk-assisted approximate algorithm for bounded np
  optimisation problems.
\newblock \emph{Quantum Information Processing}, 18\penalty0 (3):\penalty0 61,
  2019.
\newblock \doi{10.1007/s11128-019-2171-3}.

\bibitem[Wang et~al.(2018)Wang, Hadfield, Jiang, and Rieffel]{wang2018quantum}
Zhihui Wang, Stuart Hadfield, Zhang Jiang, and Eleanor~G Rieffel.
\newblock Quantum approximate optimization algorithm for maxcut: A fermionic
  view.
\newblock \emph{Physical Review A}, 97\penalty0 (2):\penalty0 022304, 2018.
\newblock \doi{10.1103/PhysRevA.97.022304}.

\bibitem[Jiang et~al.(2017)Jiang, Rieffel, and Wang]{jiang2017near}
Zhang Jiang, Eleanor~G Rieffel, and Zhihui Wang.
\newblock Near-optimal quantum circuit for grover's unstructured search using a
  transverse field.
\newblock \emph{Physical Review A}, 95\penalty0 (6):\penalty0 062317, 2017.
\newblock \doi{10.1103/PhysRevA.95.062317}.

\bibitem[Morales et~al.(2018)Morales, Tlyachev, and
  Biamonte]{morales2018variational}
Mauro~ES Morales, Timur Tlyachev, and Jacob Biamonte.
\newblock Variational learning of grover's quantum search algorithm.
\newblock \emph{Physical Review A}, 98\penalty0 (6):\penalty0 062333, 2018.
\newblock \doi{10.1103/PhysRevA.98.062333}.

\bibitem[Akshay et~al.(2020)Akshay, Philathong, Morales, and
  Biamonte]{akshay2020reachability}
V.~Akshay, H.~Philathong, M.~E.~S. Morales, and J.~D. Biamonte.
\newblock Reachability deficits in quantum approximate optimization.
\newblock \emph{Phys. Rev. Lett.}, 124:\penalty0 090504, Mar 2020.
\newblock \doi{10.1103/PhysRevLett.124.090504}.

\bibitem[Otterbach et~al.(2017)Otterbach, Manenti, Alidoust, Bestwick, Block,
  Bloom, Caldwell, Didier, Fried, Hong, et~al.]{otterbach2017unsupervised}
JS~Otterbach, R~Manenti, N~Alidoust, A~Bestwick, M~Block, B~Bloom, S~Caldwell,
  N~Didier, E~Schuyler Fried, S~Hong, et~al.
\newblock Unsupervised machine learning on a hybrid quantum computer.
\newblock \emph{arXiv preprint arXiv:1712.05771}, 2017.
\newblock URL \url{https://arxiv.org/abs/1712.05771}.

\bibitem[Willsch et~al.(2020)Willsch, Willsch, Jin, De~Raedt, and
  Michielsen]{willsch2020benchmarking}
Madita Willsch, Dennis Willsch, Fengping Jin, Hans De~Raedt, and Kristel
  Michielsen.
\newblock Benchmarking the quantum approximate optimization algorithm.
\newblock \emph{Quantum Information Processing}, 19:\penalty0 197, 2020.
\newblock \doi{10.1007/s11128-020-02692-8}.

\bibitem[Abrams et~al.(2020)Abrams, Didier, Johnson, da~Silva, and
  Ryan]{abrams2019implementation}
Deanna~M Abrams, Nicolas Didier, Blake~R Johnson, Marcus~P da~Silva, and Colm~A
  Ryan.
\newblock Implementation of xy entangling gates with a single calibrated pulse.
\newblock \emph{Nature Electronics}, 3\penalty0 (12):\penalty0 744--750, 2020.
\newblock \doi{10.1038/s41928-020-00498-1}.

\bibitem[Bengtsson et~al.(2020)Bengtsson, Vikst{\aa}l, Warren, Svensson, Gu,
  Kockum, Krantz, Kri{\v{z}}an, Shiri, Svensson, et~al.]{bengtsson2019quantum}
Andreas Bengtsson, Pontus Vikst{\aa}l, Christopher Warren, Marika Svensson, Xiu
  Gu, Anton~Frisk Kockum, Philip Krantz, Christian Kri{\v{z}}an, Daryoush
  Shiri, Ida-Maria Svensson, et~al.
\newblock Improved success probability with greater circuit depth for the
  quantum approximate optimization algorithm.
\newblock \emph{Physical Review Applied}, 14\penalty0 (3):\penalty0 034010,
  2020.
\newblock \doi{10.1103/PhysRevApplied.14.034010}.

\bibitem[Qiang et~al.(2018)Qiang, Zhou, Wang, Wilkes, Loke, O’Gara, Kling,
  Marshall, Santagati, Ralph, et~al.]{qiang2018large}
Xiaogang Qiang, Xiaoqi Zhou, Jianwei Wang, Callum~M Wilkes, Thomas Loke, Sean
  O’Gara, Laurent Kling, Graham~D Marshall, Raffaele Santagati, Timothy~C
  Ralph, et~al.
\newblock Large-scale silicon quantum photonics implementing arbitrary
  two-qubit processing.
\newblock \emph{Nature photonics}, 12\penalty0 (9):\penalty0 534--539, 2018.
\newblock \doi{10.1038/s41566-018-0236-y}.

\bibitem[Zacharov et~al.(2019)Zacharov, Arslanov, Gunin, Stefonishin, Bykov,
  Pavlov, Panarin, Maliutin, Rykovanov, and Fedorov]{Zacharov2019}
Igor Zacharov, Rinat Arslanov, Maksim Gunin, Daniil Stefonishin, Andrey Bykov,
  Sergey Pavlov, Oleg Panarin, Anton Maliutin, Sergey Rykovanov, and Maxim
  Fedorov.
\newblock {\textquotedblleft}zhores{\textquotedblright} {\textemdash} petaflops
  supercomputer for data-driven modeling, machine learning and artificial
  intelligence installed in skolkovo institute of science and technology.
\newblock \emph{Open Engineering}, 9\penalty0 (1):\penalty0 512--520, October
  2019.
\newblock \doi{10.1515/eng-2019-0059}.

\end{thebibliography}

\newpage
\onecolumngrid

\newpage

\newpage

\newpage 

\newpage 

\end{document}